\documentclass[runningheads]{llncs}
\usepackage[T1]{fontenc}
\usepackage{pdfpages}
\usepackage{ulem}
\usepackage{lineno}
\usepackage{latexsym}
\usepackage{graphicx}
\usepackage{subfigure}
\usepackage{caption}
\usepackage{amsmath}
\usepackage{multirow}
\usepackage{booktabs}
\begin{document}
\title{Contextual Dual Learning Algorithm with Listwise Distillation for Unbiased Learning to Rank}
\titlerunning{Contextual Dual Learning Algorithm with Listwise Distillation}
%
\author{Lulu Yu\inst{1,2}\and
Keping Bi\inst{1,2}\and
Shiyu Ni\inst{1,2}\and
Jiafeng Guo\inst{1,2}\thanks{Corresponding author}}

\authorrunning{Lulu Yu et al.}
%
\institute{CAS Key Lab of Network Data
Science and Technology, ICT, CAS \and
University of Chinese Academy of Sciences Beijing, China
\email{\{yululu23s,bikeping,nishiyu23z,guojiafeng\}@ict.ac.cn}
}

\maketitle              
\begin{abstract}
Unbiased Learning to Rank (ULTR) aims to leverage biased implicit user feedback (e.g., click) to optimize an unbiased ranking model. The effectiveness of the existing ULTR methods has primarily been validated on synthetic datasets. However, their performance on real-world click data remains unclear. Recently, Baidu released a large publicly available dataset of their web search logs. Subsequently, the NTCIR-17 ULTRE-2 task released a subset dataset extracted from it. We conduct experiments on commonly used or effective ULTR methods on this subset to determine whether they maintain their effectiveness. In this paper, we propose a Contextual Dual Learning Algorithm with Listwise Distillation (CDLA-LD) to simultaneously address both position bias and contextual bias. We utilize a listwise-input ranking model to obtain reconstructed feature vectors incorporating local contextual information and employ the Dual Learning Algorithm (DLA) method to jointly train this ranking model and a propensity model to address position bias. As this ranking model learns the interaction information within the documents list of the training set, to enhance the ranking model's generalization ability, we additionally train a pointwise-input ranking model to learn the listwise-input ranking model's capability for relevance judgment in a listwise manner. Extensive experiments and analysis confirm the effectiveness of our approach.

\end{abstract}
\section{Introduction}
Learning to Rank (LTR) is an important component in many real-world systems. Generally, the field of LTR utilizes human annotations as the training data where the relevance of documents is annotated by human experts. As human annotations are expensive to obtain and potentially misaligned with user preferences in the specific system, many researchers seek to leverage implicit user feedback (e.g., click). Although it can serve as an indication of search results' relevance, directly optimizing a ranking model with it suffers from a large number of biases, e.g., position bias, trust bias, and contextual bias. Thus, many Unbiased Learning to Rank (ULTR) methods have been proposed to mitigate these biases. 

Among these methods, Counterfactual Learning to Rank (CLTR) has drawn a lot of attention and has proven to be effective. The core idea is to estimate the probability of a user examining the search result. By using the inverse of the examination probability as the weight for the clicked items, an unbiased relevance estimation objective can be achieved theoretically. Thus, it is crucial to estimate the examination probability accurately. Joachims et al. \cite{joachims2017unbiased} uses the result randomization. While the result randomization hurts the user experience, Ai et al. \cite{ai2018unbiased} propose the Dual Learning Algorithm (DLA) to jointly learn an unbiased ranking model and an unbiased propensity model. Luo et al. \cite{luo2024unbiased} uncover that DLA fails to address the issue of propensity overestimation caused by the relevance confounder and propose a new learning objective based on backdoor adjustment.

Note that the above methods mainly concentrate on the position factor to tackle the position bias. However, whether a user examines the document depends not only on its position but also on the surrounding documents. Zhuang et al. \cite{zhuang2021cross} propose a cross-positional attention mechanism considering the correlation between clicks on a specific document and other documents in the same session, including their relevance and positions. Chen et al. \cite{chen2021adapting} leverage the embedding method by combining both position and click factors to develop an Interactional Observation-Based Model (IOBM) to estimate the examination probability.

Due to the lack of public datasets containing real clicks, existing ULTR methods have demonstrated their effectiveness mostly on synthetic datasets, where clicks are simulated based on the given user browsing behavior. It remains unclear whether these methods can still preserve their effectiveness in real-world datasets. Recently, Zou et al. \cite{zou2022large} publish a large-scale search dataset collected from the search logs of the largest Chinese search engine, Baidu. And for convenience, there are some subsets \cite{hager2024unbiased,niu2023overview} extracted from it. 

In this paper, firstly, we reproduce and compare some typical ULTR methods on this subset \cite{niu2023overview}, including heuristic features (e.g., BM25) and pre-trained scores produced by the pre-trained model \cite{chen2023multi}. Secondly, we propose a Contextual Dual Learning Algorithm with Listwise Distillation (CDLA-LD) to mitigate both position and contextual biases. It contains two ranking models: (1) a listwise-input ranking model leveraging the self-attention mechanism and (2) a pointwise-input ranking model. It jointly learns the unbiased listwise-input ranking model using DLA and the pointwise-input ranking model distilled from the listwise-input model in a listwise way. The listwise input combined with the self-attention mechanism can model the cross-document interactions to capture local context information in the documents list. Despite its advantages, one significant drawback is inputting the entire list of documents at once rather than directly obtaining the relevance between any query and document. The interaction mode learned in training data may be different from that in the test. To acquire a more generalized ranking model, 
we optimize another pointwise-input ranking model to distill the knowledge of the listwise-input one in a listwise way. Experiments on \cite{niu2023overview} have demonstrated our method performs the best. Moreover, we compare the propensity learned in different ULTR methods with ours.

The contributions of this paper can be summarized as follows:
\begin{enumerate}
\item[(1)] We reproduce some typical ULTR methods on one subset dataset of real-world search logs.
\item[(2)] We propose CDLA-LD to tackle position bias and contextual bias simultaneously, which jointly learns a listwise-input ranking model using DLA and a pointwise-input model distilled from the listwise-input one in a listwise manner.
\item[(3)] We conduct all these methods in real-world search logs. We verify our method's effectiveness and analyze the propensity learned by different ULTR methods.
\end{enumerate}

\section{Related Work}
\subsection{Unbiased Learning to Rank (ULTR)}
To leverage biased user feedback (e.g., click) for optimizing Learning to Rank (LTR) systems, a significant amount of research on ULTR has been proposed to acquire an unbiased ranking model. There are two primary streams for ULTR. One trend depends on click models \cite{borisov2016neural,craswell2008experimental,mao2018constructing}, which assume user browsing behavior to estimate the examination probability. By maximizing the likelihood of the observed data, unbiased relevance estimation can be accurately obtained. Reliable relevance estimation requires the same query-document pair to appear multiple times. However, this can be challenging for long-tail queries and some sparse systems, e.g., personal search. The other trend derives from counterfactual learning, which addresses bias with inverse propensity score \cite{ai2018unbiased,joachims2017unbiased,luo2024unbiased}. The key point is how to estimate propensity using bias factors appropriately. Ai et al. \cite{ai2018unbiased} propose a Dual Learning Algorithm jointly learning an unbiased ranking model and an unbiased propensity model. Subsequent methods \cite{chen2021adapting,zhuang2021cross} for improving propensity estimation can be integrated into the DLA framework.

\subsection{ULTR on the Baidu-ULTR Dataset}
Research \cite{chen2023thuir,chen2023multi,yu2023cir,yu2023feature} on Baidu-ULTR dataset \cite{zou2022large} concerning ULTR mostly revolves around two public competitions: WSDM Cup 2023 - \textit{unbiased learning-to-rank} and the ULTRE-2 task at NTCIR-17. They improve performance by combining the output of the BERT models with traditional LTR features \cite{chen2023thuir,chen2023multi,yu2023feature} or pseudo relevance feedback \cite{yu2023cir}. However, they do not extensively explore the effectiveness of ULTR methods for the competition.

Thus, in this work, we explore whether existing ULTR methods can maintain their performance on real-world click data. Moreover, we propose a Contextual Dual Learning Algorithm with Listwise Distillation (CDLA-LD), using the self-attention mechanism to model the local context information and listwise distillation to enhance the ranking model's generality capabilities.

\section{Preliminaries}
In this section, we formulate Learning to Rank (LTR) with implicit user feedback, i.e., click, and introduce the core idea of the Dual Learning Algorithm (DLA) \cite{ai2018unbiased}.
\subsection{Problem Formulation}

For a user query $q$, $\pi_q$ is the ranked list of documents for $q$. Let $d_i\in \pi_q$ be a document displayed at position $i$ and $x_i$ be the feature vector of $q-d_i$ pair. We use binary Bernoulli variables $r_i$, $e_i$, and $c_i$ to denote whether $d_i$ is relevant, examined, and clicked, respectively. LTR aims to find a function $f$ mapping from feature vector $x_i$ to relevance $r_i$, i.e., ranking model. Assuming that we already know the true relevance $r_i$ (full information setting), then we can formulate the local ranking loss $\mathcal{L}_{full-info}$ as,
\begin{equation}
    \mathcal{L}_{full-info}(f,q|\pi_q)=\sum_{d_i\in\pi_q} \Delta(f(x_i),r_i|\pi_q),
\label{full-info}
\end{equation}
where $\Delta$ is a loss function computing the individual loss on each document. When we replace the true relevance $r_i$ with the implicit user feedback, i.e., click $c_i$ in Eq.\ref{full-info}, then the empirical local ranking loss is derived as follows,
\begin{equation}
    \mathcal{L}_{naive}(f,q|\pi_q)=\sum_{d_i\in\pi_q} \Delta(f(x_i),c_i|\pi_q).
\end{equation}
As different biases exist in clicks, this naive loss function is biased. Unbiased Learning to Rank (ULTR) methods aim to learn an unbiased ranking model $f$ with these biased clicks.

\subsection{Dual Learning Algorithm (DLA)}
Based on the examination hypothesis,
\begin{equation}
    p(c_i=1|\pi_q)=p(r_i=1|\pi_q)·p(e_i=1|\pi_q),
\label{eh}
\end{equation}
Joachims et al. \cite{joachims2017unbiased} propose the Inverse Propensity Weighting (IPW) method to acquire an unbiased ranking objective. According to Eq.\ref{eh}, Ai et al. \cite{ai2018unbiased} discover that the problem of estimating examination propensity is symmetric with the problem of estimating real document relevance from user clicks. Thus, they propose the Dual Learning Algorithm (DLA) to solve the two problems simultaneously. Specifically, in DLA, an unbiased ranking model $f$ and an unbiased propensity model $g$ can be jointly learned by optimizing the following local loss,
\begin{equation}
    \begin{aligned}
        &\mathcal{L}_{IPW}(f,q|\pi_q) = \sum_{d_i\in\pi_q, c_i=1}\frac{\Delta (f(x_i), c_i|\pi_q)}{g(i)}, \\
        &\mathcal{L}_{IRW}(g,q|\pi_q) = \sum_{d_i\in\pi_q, c_i=1}\frac{\Delta (g(i), c_i|\pi_q)}{f(x_i)},
    \end{aligned}
\end{equation}
where $g$ is the function of position $i$ as DLA merely considers the position factor to estimate the position bias.

DLA employs listwise loss and normalizes the inverse probability weights. In fact, the two inverse probability weighted loss functions in DLA are formulated as follows,
\begin{equation}
    \begin{aligned}
        &\mathcal{L}_{IPW}(f,q|\pi_q) = -\sum_{d_i\in\pi_q, c_i=1} \frac{g(1)}{g(i)}·log\frac{e^{f(x_i)}}{\sum_{z_j\in\pi_q}e^{f(x_j)}}, \\
        &\mathcal{L}_{IRW}(g,q|\pi_q) = -\sum_{d_i\in\pi_q, c_i=1} \frac{f(x_1)}{f(x_i)}log\frac{e^{g(i)}}{\sum_{z_j\in\pi_q}e^{g(j)}},
    \end{aligned}
    \label{dla_loss}
\end{equation}

\section{Contextual Dual Learning Algorithm with Listwise Distillation (CDLA-LD)}
Users often interact with relevant documents from a list rather than independently evaluate each document. Thus, modeling this interaction and obtaining the reconstructed document representation containing local context information is ignorant. Utilizing more representative vectors may estimate the position bias more accurately and realistically.

\subsection{Overview}

\begin{figure}
    \centering
    \includegraphics[width=0.8\textwidth]{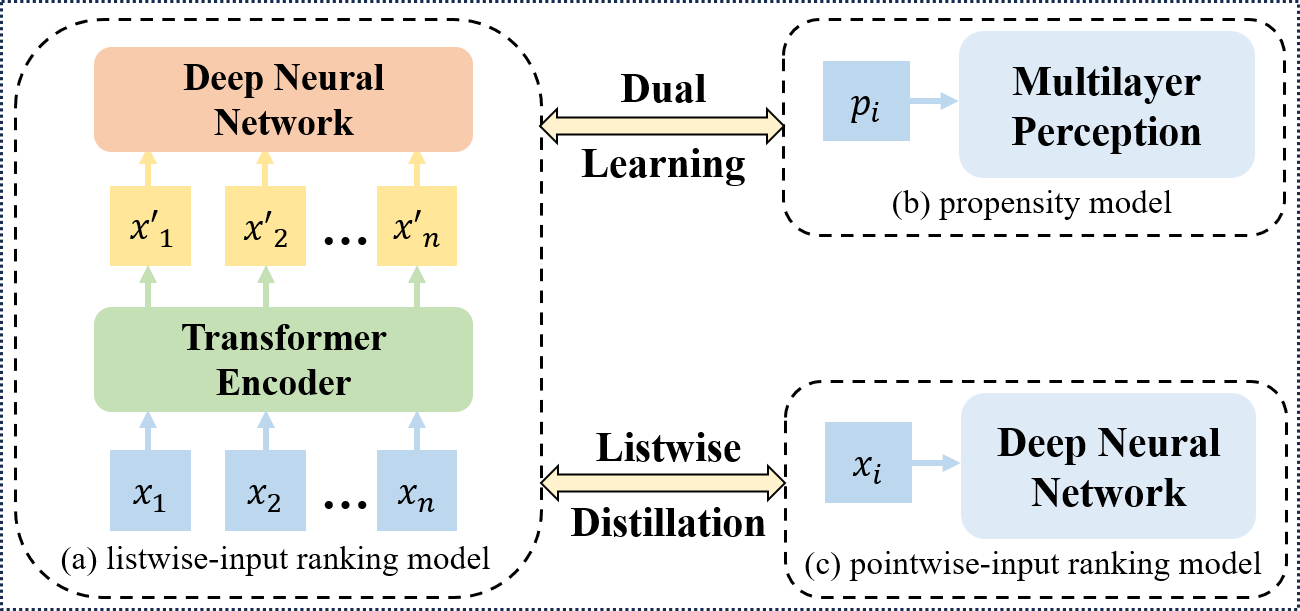}
    \caption{The overall architecture of CDLA-LD, where we jointly train an unbiased listwise-input ranking model (a) employing Transformer Encoder and an unbiased propensity model (b). And simultaneously we learn a pointwise-input ranking model (c) distilled from (a) in a listwise manner.}
    \label{architecture}
\end{figure}

The overall architecture of our proposed CDLA-LD is shown in Fig. \ref{architecture} where $n$ represents the list length, 10 in our work. It mainly consists of two procedures: the Contextual Dual Learning Algorithm (CDLA) and the Listwise Distillation (LD). 

Taking the documents list as the input of a listwise-input ranking model, it first passes through a transformer encoder module to obtain document vectors integrated with contextual information. These vectors are then used to estimate relevance. Similar to DLA, to address position bias, this ranking model is learned jointly with the propensity model. Concurrently, we train a pointwise-input ranking model distilling the knowledge learned from the listwise-input one. 

\subsection{Contextual Dual Learning Algorithm (CDLA)}
Inspired by the self-attention mechanism \cite{parikh2016decomposable}, which integrates the context information by computing a multiple-weighted sum of the vectors in the current sequence. Therefore, we utilize the Transformer Encoder \cite{vaswani2017attention} to obtain context-included vectors and then generate the relevance estimation through the Deep Neural Model (DNN). These two parts, Transformer Encoder and DNN constitute the listwise-input ranking model. 

We directly employ the Dual Learning Algorithm (DLA) \cite{ai2018unbiased} to address the position bias. We also consider that the examination probability solely depends on the position, unlike \cite{chen2021adapting} considering both position and click factors. Thus, we jointly optimize the listwise-input ranking model and the propensity model using Eq. \ref{dla_loss}. 

\subsection{Listwise Distillation}
The Transformer Encoder component in the listwise-input ranking model models the interaction information among documents and, through training, learns the implicit interaction patterns among documents in the training set. During the inference stage, on the one hand, interactions within its lists may differ from those present in the training set, thereby compromising the generalization ability of the ranking model. On the other hand, introducing the Transformer Encoder structure could increase computational overhead, thus reducing inference efficiency.

To enhance the generalization ability and improve inference efficiency, inspired by knowledge distillation, we introduce a pointwise-input ranking model to learn the relevance judgment capabilities of the listwise-input ranking model in a listwise manner. We utilize the local listwise distillation loss as follows to train a pointwise-input ranking model,
\begin{equation}
\mathcal{L}_{distillation}=-\sum_{d_i\in\pi_q}\frac{e^{f(x_i)}}{\sum_{z_j\in\pi_q}e^{f(x_j)}}log\frac{e^{h(x_i|\pi_q)}}{\sum_{z_j\in\pi_q}e^{h(x_j|\pi_q)}},
\end{equation}
where $h$ represents the pointwise-input ranking model.

The pointwise-input ranking model (i.e., student model) is employed for relevance evaluation, while the listwise-input ranking model serves as an intermediate auxiliary model (i.e., teacher model).

Note that due to the inability to accurately assess the ranking performance of the listwise-input ranking model, the training of the two ranking models is conducted simultaneously.

\section{Experiments}
We present our experimental settings and empirical results, substantiating the effectiveness of our method through the investigation of the following research questions (RQs):

\begin{itemize}
\item[$\bullet$]\textbf{RQ1:} How does CDLA-LD perform compared to other Unbiased Learning to Rank (ULTR) methods on real-world click data?
\item[$\bullet$]\textbf{RQ2:} Is the propensity estimation of each position by CDLA-LD more accurate compared to other ULTR methods?
\item[$\bullet$]\textbf{RQ3:} Is it effective for CDLA-LD to conduct debiasing when training the listwise-input ranking model?
\item[$\bullet$]\textbf{RQ4:} Is each component of CDLA-LD necessary?
\end{itemize}

\subsection{Experimental Settings}
\textbf{Datasets.} We conduct a series of experiments on a subset \cite{niu2023overview} of the Baidu-ULTR dataset \cite{zou2022large}, the first billion-level dataset for ULTR collected from the mobile web search engine of Baidu. This subset involves $1,052,295$ searching sessions ($34,047$ unique queries) with click information and $7,008$ expert annotated queries with the frequency of queries, where each query-document pair is annotated with 5-level labels.

Following \cite{zou2022large}, the expert annotation dataset is
split into validation and test sets according to a $20\%-80\%$ scheme. For training data, we remove the sessions with less than 10 consecutively recorded candidate documents, as well as those without any clicks at the top 10 position. In addition, we also split the left sessions into train, validation, and test sets according to a $80\%-10\%-10\%$ scheme. The final training data contains $485,342$ sessions ($33,302$ unique queries).

This subset provides us with three types of features, i.e., heuristic features (e.g., TF-IDF, BM25), as well as a score feature and 768 dimensions' embedding features extracted from a pre-trained model\footnote{\url{ https://github.com/lixsh6/Tencent\_wsdm\_cup2023}}. In our work, we choose the first two types of features (i.e., 14 numerical features), wherein directly using the pre-trained score is equivalent to freezing the parameters of the pre-trained model during the training stage.

\noindent\textbf{Baselines.} We compare our CDLA-LD with the following baselines, which are widely used or effective. Note that for the sake of comparability, we apply listwise loss to all the methods described below.
\begin{itemize}
\item[$\bullet$]\textbf{Naive:} It directly trains the model with clicks without debiasing.
\item[$\bullet$]\textbf{IPW:} Inverse Propensity Weighting is one of the first ULTR algorithms proposed under the framework of counterfactual learning \cite{joachims2017unbiased}, which weights the training loss with the probability of the document being examined in the search session.
\item[$\bullet$]\textbf{DLA:} The Dual Learning Algorithm \cite{ai2018unbiased} simultaneously learns an unbiased ranking model and an unbiased propensity model.
\item[$\bullet$]\textbf{XPA:} Zhuang et al. \cite{zhuang2021cross} propose Cross-Positional Attention to effectively capture cross-positional interactions among displayed documents.
\item[$\bullet$]\textbf{UPE:} Luo et al. \cite{luo2024unbiased} propose Unconfounded Propensity Estimation to address the propensity overestimation issue based on backdoor adjustment.
\item[$\bullet$]\textbf{IBOM-DLA:} Chen et al. \cite{chen2021adapting} leverage the embedding method to develop an Interactional Observation-Based Model (IOBM) to estimate the examination probability, here we integrate the IOBM with DLA.
\end{itemize}

\noindent\textbf{Evaluation Metrics.} To assess the performance of ranking, we use the normalized Discount Cumulative Gain (nDCG) \cite{jarvelin2002cumulated} and Expected Reciprocal Rank (ERR) \cite{chapelle2009expected}. For both metrics, we report the results at rank 1, 3, 5, 10.

\noindent\textbf{Implementation Details.} We implement our CDLA-LD and utilize the baselines in the ULTRA framework \cite{tran2021ultra} to conduct our experiments. We use AdamW \cite{loshchilov2017decoupled} to train all the models with a learning rate tuned in the range of 2e-6 to 2e-5. We set the batch size to 30 (the number of queries) and fix the length of the documents list in the training stage as 10.

We tune the Transformer Encoder architecture within the listwise-input ranking model. Specifically, we tune the number of layers $\in \{1, 2, 3, 4\}$, the number of heads $\in \{2, 4, 8\}$ on the validation. 

For IPW, following \cite{hager2024unbiased}, the propensity is estimated by the AllPairs intervention harvesting method \cite{agarwal2019estimating}.

The Deep Neural Network (DNN) component in both the listwise-input ranking model and the pointwise-input ranking model is identical, with three hidden layers with sizes $[32, 16, 8]$, with the ELU \cite{clevert_fast_2016} activation function. Additionally, in both ranking models, the feature vectors are first passed through a linear layer to be expanded to 64 dimensions from 14 dimensions. 

\subsection{Performance Comparison of Different ULTR Methods (RQ1)}
Tab. \ref{overall_comp} shows the overall performance comparison of different ULTR methods. In contrast to the Naive, IPW even yields inferior performance, and DLA does not demonstrate significant improvement, while XPA and IBOM-DLA, which introduce complex information to model the propensity model, show different degrees of improvement on nDCG. The UPE method, designed to address the propensity overestimation, also shows improvements in most metrics compared to DLA. These results once again highlight the impact of propensity estimation on the performance of ranking models. 

Unlike XPA and IBOM, which modify the propensity model's modeling, our CDLA-LD considers implicit interactions between documents to obtain reconstructed feature vectors of documents that contain local context information. It shows varying degrees of enhancement across all the metrics compared to all the baselines.

\begin{table}
\centering
\caption{Performance comparison between CDLA-LD and different baselines. The method in \textbf{bold} has the best performance, while the method \uline{underlined} has the second-best performance. "\dag" and "\ddag" indicate statistically significant improvements (t-test with p-value $\leq$ 0.05) over DLA and the best baseline, respectively. }
\label{overall_comp}
\begin{tabular}{ccccccccc}
\hline
\multirow{2}*{Methods} & \multicolumn{4}{c}{nDCG@K} & \multicolumn{4}{c}{ERR@K} \\
\cmidrule(r){2-5}\cmidrule(r){6-9}
~ & K=1 &  K=3 &  K=5 &  K=10 & K=1 &  K=3 &  K=5 &  K=10 \\
\hline
Naive & 0.4407 & 0.4534 & 0.4669 & 0.4967 & 0.1529 & 0.2364 & 0.2637 & 0.2845 \\
IPW & 0.3985 & 0.4197 & 0.4350 & 0.4675 & 0.1432 & 0.2232 & 0.2497 & 0.2704 \\
DLA & 0.4417 & 0.4546 & 0.4675 & 0.4975 & 0.1533 & 0.2369 & 0.2641 & 0.2849 \\
XPA & 0.4430 & 0.4552 & 0.4680 & 0.4981 & 0.1519 & 0.2358 & 0.2632 & 0.2839 \\
UPE & 0.4387 & 0.4553 & 0.4678 & 0.4998 & 0.1547 & 0.2387 & 0.2656 & 0.2865 \\
\uline{IBOM-DLA} & \uline{0.4450} & $\uline{0.4590}^\dag$ & $\uline{0.4718}^\dag$ & $\uline{0.5024}^\dag$ & $\uline{0.1550}^\dag$ & $\uline{0.2398}^\dag$ & $\uline{0.2668}^\dag$ & $\uline{0.2876}^\dag$ \\
\textbf{CDLA-LC} & $\textbf{0.4469}^\dag$ & $\textbf{0.4614}^\dag$ & $\textbf{0.4733}^\dag$ & $\textbf{0.5025}^\dag$ & $\textbf{0.1576}^{\dag\ddag}$ & $\textbf{0.2418}^{\dag\ddag}$ & $\textbf{0.2686}^\dag$ & $\textbf{0.2890}^\dag$ \\
\hline
\end{tabular}
\end{table}

\subsection{Propensity Analysis (RQ2)}
\begin{figure}
    \centering
    \subfigure[]{\includegraphics[width=0.4\textwidth]{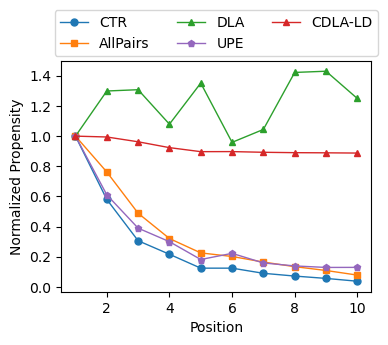}
    \label{prop}
    \captionsetup{justification=raggedright, singlelinecheck=false}}
    \subfigure[]{\includegraphics[width=0.4\textwidth]{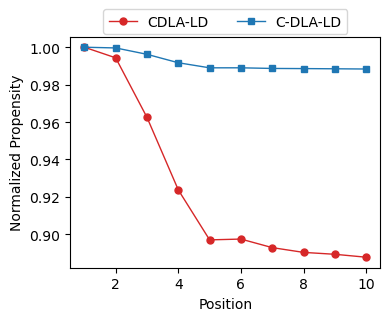}
    \label{time_prop}
    \captionsetup{justification=raggedleft,singlelinecheck=false}}
    \caption{(a) Normalized propensity estimated from ULTR methods that solely consider position factor to model the propensity model, compared to the mean CTR. (b) Normalized propensity estimated from CDLA-LD and C-DLA-LD.}
\end{figure}

Fig. \ref{prop} displays normalized propensity ($\frac{g(i)}{g(1)}$) estimated from AllPairs, DLA, UPE, CDLA-LD, and the click-through rate (CTR) per position. The propensity learned by DLA is somewhat chaotic. The propensity learned by UPE and AllPairs is close to CTR. Nevertheless, the performance of the IPW method using propensity estimated by AllPairs is significantly lower compared to UPE. This may be because, under UPE, the ranking and propensity models are trained simultaneously rather than fixing the propensity for training the ranking model.

CDLA-LD learns that the propensity decreases gradually with increasing position. The specific normalized propensity for each position is as follows: 
\textbf{1.000}, \textbf{0.994}, 0.962, 0.924, 0.897, 0.897, 0.893, 0.890, 0.889, 0.888. This propensity closely reflects the actual likelihood of users browsing each position in real-world scenarios, especially in mobile search engines. For instance, the examination probability for the first two positions is approximately equal, as the first two positions are usually seen simultaneously after the documents list is returned.

\subsection{Study on When to Debias (RQ3)}
Our approach, CDLA-LD, involves training an unbiased listwise-input ranking model and distilling it so that a pointwise-input ranking model learns the implicit relevance judgment capability of the listwise-input ranking model. Similarly, one can consider fitting click signals with the listwise-input model, whose outputs represent the click probability, and subsequently training an unbiased pointwise-input ranking model based on it. We refer to the latter as C-DLA-LD in short, where C and LD denote Contextual and Listwise Distillation, respectively.

Tab. \ref{time_perf} presents the performance comparison between CDLA-LD and C-DLA-LC, while Fig. \ref{time_prop} illustrates the learned normalized propensity of both methods. The debiasing effect is pronounced when training the listwise-input ranking model. The debiasing effect during the training of the pointwise-input ranking model outperforms DLA, which may reflect the advantages of the listwise-input model. Therefore, training an unbiased listwise-input ranking model and distilling it into a pointwise-input ranking model can achieve better performance.

\begin{table}
\centering
\caption{Performance comparison between CDLA-LD and C-DLA-LD. The method in \textbf{bold} has the best performance. "*" indicates statistically significant improvements (t-test with p-value $\leq$ 0.05) over the other method.}
\label{time_perf}
\begin{tabular}{ccccccccc}
\hline
\multirow{2}*{Methods} & \multicolumn{4}{c}{nDCG@K} & \multicolumn{4}{c}{ERR@K} \\
\cmidrule(r){2-5}\cmidrule(r){6-9}
~ & K=1 & K=3 &  K=5 &  K=10 & K=1 & K=3 &  K=5 &  K=10 \\
\hline
{C-DLA-LD} & 0.4385 & 0.4562 & 0.4694 & 0.4990 & 0.1547 & 0.2390 & 0.2659 & 0.2864 \\
\textbf{CDLA-LD} & $\textbf{0.4469}^*$ & $\textbf{0.4614}^*$ & $\textbf{0.4733}^*$ & $\textbf{0.5025}^*$ & $\textbf{0.1576}^*$ & $\textbf{0.2418}^*$ & $\textbf{0.2686}^*$ & $\textbf{0.2890}^*$ \\
\hline
\end{tabular}
\end{table}


\subsection{Ablation Study (RQ4)}
Compared to DLA, CDLA-LD initially introduces a listwise-input ranking model and subsequently incorporates listwise distillation. Thus, we solely evaluate the performance of the listwise-input ranking model itself on the test set, excluding listwise distillation.

From Tab. \ref{ablation_perf}, the performance of the listwise-input ranking model on the test set significantly lags behind that of CDLA-LD. This indicates discrepancies in the interaction of document lists between the test and training sets. On the one hand, the number of documents on the list for
training and testing can seldom be exactly aligned, while on the other hand, it may stem from inherent inconsistencies in document relevance distribution.

\begin{table}
\centering
\caption{Performance comparison between CDLA-LD and CDLA. The method in \textbf{bold} has the best performance. "*" indicates statistically significant improvements (t-test with p-value $\leq$ 0.05) over the other method.}
\label{ablation_perf}
\begin{tabular}{ccccccccc}
\hline
\multirow{2}*{Methods} & \multicolumn{4}{c}{nDCG@K} & \multicolumn{4}{c}{ERR@K} \\
\cmidrule(r){2-5}\cmidrule(r){6-9}
~ & K=1 &  K=3 &  K=5 &  K=10 & K=1 &  K=3 &  K=5 &  K=10 \\
\hline
{CDLA} & 0.4157 & 0.4323 & 0.4472 & 0.4769 & 0.1492 & 0.2305 & 0.2574 & 0.2776 \\
\textbf{CDLA-LD} & $\textbf{0.4469}^*$ & $\textbf{0.4614}^*$ & $\textbf{0.4733}^*$ & $\textbf{0.5025}^*$ & $\textbf{0.1576}^*$ & $\textbf{0.2418}^*$ & $\textbf{0.2686}^*$ & $\textbf{0.2890}^*$ \\
\hline
\end{tabular}
\end{table}

\section{Conclusion}
In this work, we propose a Contextual Dual Learning Algorithm with Listwise Distillation (CDLA-LD) to mitigate both position and contextual biases. Specifically, we jointly train an unbiased listwise-input ranking model and an unbiased propensity model to address position bias. The listwise-input ranking model uses a Transformer Encoder, which can model the cross-document interactions and then capture local context information within the documents list. Considering the data mismatching between the training set and the test set, in order to enhance the generalization capability of the ranking model, we introduce a pointwise-input ranking model to distill the ability to estimate relevance from the listwise-input ranking model in a listwise manner.

Compared to other effective ULTR methods, our approach performs the best on real-world click data collected from the user logs in the Baidu web search engine. Furthermore, the propensity estimation learned by CDLA-LD intuitively aligns closely with real-world scenarios. We also validate the necessity of introducing listwise distillation.

In subsequent work, we may consider incorporating interaction information derived from the position and click of documents within document lists when modeling the propensity model.

%
%
\bibliographystyle{splncs04.bst}
\bibliography{reference}

\end{document}